# Photoluminescence properties of forsterite ($Mg_2SiO_4$) nanobelts synthesized from Mg and $SiO_2$ powders

R. Zhu[*], X. S. Peng, S. H Sun, Y. Lin, L. D. Zhang,

Institute of Solid State Physics, Chinese Academy of Sciences, Hefei 230031,

P. R. China

**Abstract**

Ternary oxide forsterite $Mg_2SiO_4$ single-crystalline nanobelts have been synthesized at high yields via $SiO_2$ assistant-oxide through a VS process. The PL properties of the forsterite $Mg_2SiO_4$ nanobelts were observed. The emission peak at 625 nm is attributed to the $^4T_1 \to {}^6A_1$ transition of the trace $Mn^{2+}$ ions. The other emission peak at 400nm may be due to the defects.

---

[*] To whom correspondence should be addressed : Fax: +86-551-5591434

Email: zhurui@issp.ac.cn

# 1 Introduction

Nano-scaled materials are widely recognized as the key materials in nano-fabrication. They play decisive roles in nano-technology due to their nanometer-sized geometry and unique properties.[1-4] The fundamental characteristics of the nanostructure materials arise from the interplay of the electronic structure and the geormotry.[5] In order to elucidate the role of the interplay, it is necessary at first to prepare nano-scaled materials of any desired size and geometry.[6-9] Aside from nanotubes,[4] various of nanowires (nanorods) have been fabricated by using different methods such as laser ablation,[9] templating,[10] arc discharge,[11] vapor-phase transport,[12] and solution method.[13] Much attention has been paid to synthesize binary oxide nanobelts such as ZnO, CdO, $Ga_2O_3$ and $SnO_2$ by simple vapor-phase deposition from oxide sources.[14] However, the synthesis of ternary oxide nanobelts have not been reported previously.

Ternary oxide forsterite (fo-magnesium orthosilicate, $Mg_2SiO_4$), the magnesium end member of the olivine group of minerals, consists of $SiO_4$ tetrahedral linked by magnesium actions in octahedral coordination. It is of well-known scientific and technological interest because its physical and chemical properties in the primordial solar nebulae are believed to play an important role in the fractionation of elements prior to planet formation.[15-16] In addition, forsterite is a major component of the earth's upper mantle and is thus of considerable importance in controlling its seismic properties, rheological behavior and thermal structure.[17] Moreover, with discovery of the near-infrared lazing properties of chromium-doped signal crystals, there has recently been renewed interest in forsterite in materials science.[18-19] Conventionally, forsterite can be synthesized by solid-state reactions between MgO and $SiO_2$ at elevated temperatures (1100-1400 $^O$C).[20, 21] However, this requires high temperatures, thus several sol-gel based routes have been used to synthesize forsterite powders, often with trace enstatite.[22-24] In this communication, we have successfully synthesized single crystalline forsterite nanobelts by chemical vapor deposition from Mg powders and $SiO_2$ nanoparticles. The very interesting thing is that we have observed photoluminescence emission from these $Mg_2SiO_4$ nanobelts.

**2 Experiment Section**

**2. 1 Example preparation**

A mixture of Mg powders (99.99%, <0.5 μm) and $SiO_2$ nanoparticle powders (about 80 nm) was placed on a ceramic boat. This boat was put into a ceramic tube (25 mm inner diameter) that was inserted in a horizontal electronic resistance furnace. Then the furnace was rapidly heated up to 1100 °C, and kept at this temperature for 60 min with constant flow of argon gas at a flow rate of 200 sccm (standard cubic centimeters/minute). After the system had cooled down to room temperature, a large piece of white, wool-like product was collected from the inner wall of the ceramic boat at the downstream end.

**2.2 Characterization and PL, EPR Measurements**

The products were characterized by X-ray diffraction (XRD) (PW 1710 instrument with Cu Kα radiation, operated at 40 keV and 40 mA), scanning electron microscopy (SEM) (JEOL JSM 6300F), high-resolution transmission electron microscopy (HRTEM) (JEOL 2010, operated at 200 kV), Energy–dispersive X-ray fluorescence (EDX) (EDAX, DX-4) attached to the JEOL 2010. For SEM observations, the product was pasted on the Al substrate by carbon conducting paste. Specimens for TEM and HRTEM investigation were briefly ultra-sonicated in ethanol, and then a drop of suspension was placed on a holey copper grid with carbon film. The electron paramagnetic resonance (EPR) spectrum was carried out by using a Bruker ER-200D electron paramagnetic resonance device with 10 dB (20 mw) at room temperature. Photoluminescence spectra were obtained using a Hitachi 850-fluorescence spectrophotometer with Xe lamp at room temperature.

## 3 Result and Discussion

The phase purity of the as-synthesized products was examined by XRD. All of the reflection of the XRD pattern in Fig.1 can be indexed to a pure orthorhombic phase [space group: Pbnm (62)] of forsterite $Mg_2SiO_4$ with lattice constants of a = 4.755 Å, b = 10.198 Å and c = 5.979 Å (JCPDS 84-1402).

The morphology, structure and composition of the as-synthesized products have been characterized using SEM, HRTEM and EDS. Typical SEM images of the products are shown in Fig.2. A low-magnification SEM (Fig.2a) shows some straight and twisted wire-like nanostructures with several to hundreds micrometers long. A representative high-magnification SEM image (Fig. 2b) of two belts reveals that each nanobelt has a uniform width about several hundreds nanometers and thickness about tens nanometers along its entire length, displaying the shape characteristics of the belts. The typical width of the nanobelts are in the range of 100-500 nm, the thickness and width-to-thickness ratios of the $Mg_2SiO_4$ nanobelts are in the range of 10-30 nm and ~10-50, respectively, as determined from several tens of nanobelts.

Fig. 3a and b show the $Mg_2SiO_4$ nanobelts were synthesized at 1100 $^O$C for 30 and 60 min, respectively. It can be seen that the geometrical shape of the as-synthesized nanostructures is a belt that is distinct in cross-section from the previously reported nanotubes[4] and nanowires.[9-13] Each nanobelt has a uniform width and thickness along its entire length, and no nanoparticles were observed at the tips of the nanobelts. A ripple-like contrast observed in TEM images is due to strain from the bending of the belt. In addition, it is clearly that the nanobelt (about 150 nm in width) shown in Fig. 3a is smaller than that (about 500 nm in width) shown in Fig.3b. This result demonstrates that if the reaction time is shorter, the dimensions of the forsterite nanobelts will be smaller. The SAED pattern and HRTEM image (Fig.3c) reveal that the forsterite nanobelts are structurally uniform and single crystalline. The SAED pattern (Fig.3c, inset) recorded perpendicular to the nanobelt long axis could be indexed for the [010] zone axis of single crystalline forsterite $Mg_2SiO_4$ and suggests that the nanobelt growth along the <101> direction. Fig.3c also shows a lattice–resolution HRTEM of the nanobelt with 0.476 nm and 0.598 nm lattice fringes in agreement with the d values of the (100) and (001) planes of the forsterite crystal. The surfaces of the nanobelts are clean, atomically sharp and without any sheathed amorphous phase. In addition, the HRTEM image and SAED pattern show that the flat plane of the nanobelts is {010} surface. This non-dipolar termination of {010} surface, which is a closed-packed plane with a small surface area (28.6 $Å^2$ per unit cell), with small relaxed surface energy of 1.28 J $m^2$, is the major cleavage plane of forsterite and the most stable surface under dry conditions,[20] which is introduced to minimize the surface area possibly for reducing the total surface energy. Therefore, these forsterite nanobelts with flat plane {010} are more stable than with other flat planes. EDX analysis (Fig.3d) of this nanobelt reveals that the compositions of the nanobelt are Mg 28.4 at%, and Si 14.2 at% and O 57.4 at% similar to that of bulk $Mg_2SiO_4$ (Mg: Si:O=2:1:4).

Fig. 4 shows the PL spectrum of the forsterite nanobelts. The peak shape is distinctive in having two pronounced sub- peaks. The spectrum is disconvoluted into ultraviolet peak 1 at 3.1 eV (about 400 nm) and red peak 2 at 2.0 eV (about 627 nm) by best fitting with Gassian functions. As we know that bulk forsterite cannot emit light at room temperature. However, the present measurement (Fig. 4) shows a strong and broad PL emission spectrum from the forsterite nanobelts.

What are the origins of these PL peaks? Fig. 5 shows the EPR spectrum of the as-synthesized products measured at room temperature. The well resolved 6-fold structure (marked A) is consistent with $Mn^{2+}$ (nuclear spin I=5/2 with 100% abundance), and it matches well with that reported in references[25-27]. This result confirms that trace $Mn^{2+}$ ions exist in our products, which possibly come from the impurity Mn in the Mg powders or from the ceramic tube or boat. However, these $Mn^{2+}$ ions didn't detected by the EDS. The reason may be that the content of $Mn^{2+}$ ions is too little to be detected due to the limitation sensitivities of EDS. From Fig. 4, the EPR parameters for the $Mn^{2+}$ center are obtained as $g$ = 2.001 and hyperfine constant $A$ = 86.6 Gauss similar to that for the $Mn^{2+}$ in Mn-doped MgO.[28] However, the structure (marked B) in Fig. 4, which can not be indexed to $Mn^{2+}$ ions, indicates that some other defects may exist in the forsterite nanobelts. It has been reported that there existed two broad bands centered at 740 and 460 nm, which attributed to the $^4A_2 \rightarrow {}^4T_2$ and $^4A_2 \rightarrow {}^4T_1$ absorptive transitions in $Cr^{3+}$-doped forsterite.[18] This result indicates that impurity-doped forsterite may yield new bands. Therefore, the trace $Mn^{2+}$ ions may generate new energy levels in our product. Recently, many work have been reported on the PL properties of $Mn^{2+}$- doped nanoparticles, resulting that the emission band at about 600 nm is mainly due to the $^6A_1 \leftarrow {}^4T_1$ transition of $Mn^{2+}$.[25-27] In our case, the peak at 625 nm may be attributed to this transition. As for the red shift, it may be due to the dimension confinement (the band energy is often blue shift from larger size to smaller size nanomaterials). The average dimensions of the forsterite nanobelts (30 nm in thick, 300 nm in width and 100 μm in length) are lager than the diameters of the Mn:ZnS or Mn:CdS nanoparticles (several nm). [25-27]

The emission at 400 nm may be related to the defect centers, which were confirmed by the EPR structure B. In our case, during the rapid

evaporation–oxidation process of forsterite nanobelts, oxygen vacancies or other defects, such as Mg-Si partial disorder, local charge unbalance[29, 30] would be formed because of partially incomplete oxidation or other effects. In addition, the forsterite nanobelts with high aspect ratio and peculiar morphologies should also favor the existence of oxygen vacancies or other defects on the surface. These defects will represent new energy levels in the band gap and possibly respond for the emission.[31, 32] From Fig. 4, it can be seen that the emission intensities of the peak at 625 nm is larger than that of the peak at 400 nm. This result indicates the number of $Mn^{2+}$ ions is larger than that of defects.

As for the growth mechanism of the forsterite nanobelts, we find that there are no spherical droplets, which are known to be good evidence for vapor-liquid-solid (VLS)[9] growth mechanism, at the tips of these $Mg_2SiO_4$ nanobelts. This observation quantitatively suggests that the nanobelts do not grow by VLS mechanism proposed for nanowires grown by a catalytic technique.[9] It is likely that the $Mg_2SiO_4$ nanobelts follow a growth mechanism similar to the vapor-solid (VS) mechanism for the binary oxide nanobelts.[14] To understand the growth of these $Mg_2SiO_4$ nanobelts, the following reactions should be taken into account.

$$Mg(g)+SiO_2(s)----MgO(s)+SiO(g) \quad\quad (1)$$

$$2Mg(g)+4SiO(g)-----Mg_2SiO_4(s)+3Si(s,l) \quad\quad (2)$$

$$Si(s,l)+SiO_2(s)----2SiO(g) \quad\quad (3)$$

$$2MgO(s)+SiO_2(s)----Mg_2SiO_4(s) \quad\quad (4)$$

The melting and boiling points of Mg are 651 and 1107 °C, that of Si are 1410 and 2355 °C, that of MgO are 2800 and 3600 °C, and $SiO_2$ are 1610 and 2950 °C respectively. The melting point of $Mg_2SiO_4$ is 1890 °C. SiO is vapor phase at the temperature above 930 °C.[33] In addition SiO vapor can be formed due to the existence of an eutectic in the Si-SiO$_2$ system.[34] Therefore, at 1100 $^{O}$C temperature used in our experiment, gas species SiO(g) and Mg(g) can be continuously originated. SiO(g) is produced by reactions: (Eq.1) between gaseous Mg and SiO$_2$ nanoparticles or can appear as a product of reaction (Eq.3) between the yield Si and SiO$_2$ nanoparticles. Then Mg$_2$SiO$_4$ can be produced by reaction (Eq.2) between Mg(g) and SiO(g). There have been reported that Mg$_2$SiO$_4$ can be formed by solid reaction (Eq. 4) between MgO and SiO$_2$ at 850 or 1100 $^{O}$C.[20, 21] Therefore the solid reaction (Eq.4) will also take place. But the reaction (Eq.2) will be preferred due to the gaseous nature of this process. During the reaction progress, a little yellow smoke was continuously flowing out from the downstream end of the ceramic tube. This smoke was detected as SiO gas. The leftover of the reaction mixture was SiO$_2$ and some Mg$_2$SiO$_4$ without MgO, because SiO$_2$ was excess in the source materials. We have found that if the reaction time is longer, more Mg(g), SiO(g) and MgO(s) are generated resulting in forming thicker Mg$_2$SiO$_4$ nanobelts. So thinner Mg$_2$SiO$_4$ nanobelts can be obtained through a shorter reaction time.

**4 Conclusion**

In summary, ternary oxide forsterite Mg$_2$SiO$_4$ nanobelts with uniform width, thickness, and length have been synthesized at high yields via SiO$_2$ assistant-oxide through a VS process. The PL properties of the forsterite Mg$_2$SiO$_4$ nanobelts were observed. The emission peak at 625 nm is attributed to the $^4T_1 \rightarrow {^6A_1}$ transition of the trace Mn$^{2+}$ ions. The other emission peak at 400nm may be due to the defects in the nanobelt. With a well-defined geometry and perfect crystal structure, the nanobelts are likely to be a model materials family for a systematic understanding in the electronical, thermal, optical and ionic transport processes as well as mechanical behavior 1D nanostructures with the absence of dislocations. This will be an exciting new field for nanomaterials.

The synthesis route of forsterite $Mg_2SiO_4$ nanobelts, $SiO_2$ assistant-oxide process will offer great opportunity for the scale up preparation of other metal oxide $M_xO_y$ or ternary oxide $M_ySiO_x$ 1D nanostructure materials (M is the element who will be oxided by $SiO_2$), such as $Al_2O_3$ nanowires and nanobelts, $SnO_2$ nanobelts, $SiO_x$ nanostructures.

**Acknowledgment**

This work was supported by the Ministry of Science and Technology of China (Grant G1999064501) and natural Science Foundation of China (Grant 19974055).

**Captions for figures**

Fig. 1 The XRD pattern of the as-synthesized product.

Fig. 2 (a) A low-magnification SEM image and (b) a high-magnification SEM image of the product.

Fig.3 (a) and (b) show the TEM images of the $Mg_2SiO_4$ nanobelts synthesized at 1100 $^O$C for 30 min and 60 min, respectively. (c) HRTEM images of the products; the inset is the SAED pattern. (d) EDX analysis of the nanobelt.

Fig. 4 PL spectrum of the product.

Fig. 5 EPR spectrum of the as-synthesized products measured at room temperature.

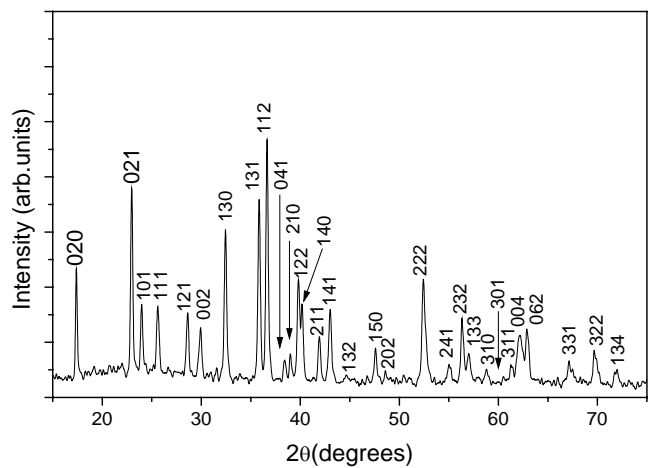

**Fig.1 R. Zhu, et al.**

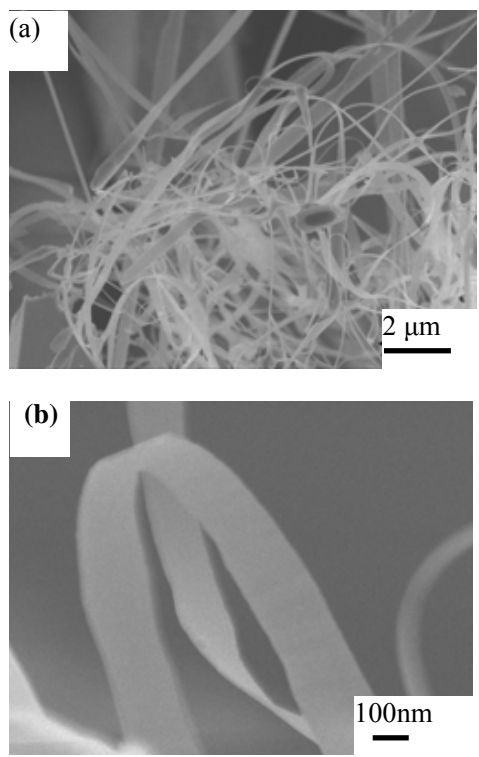

**Fig. 2 R. Zhu, et al.**

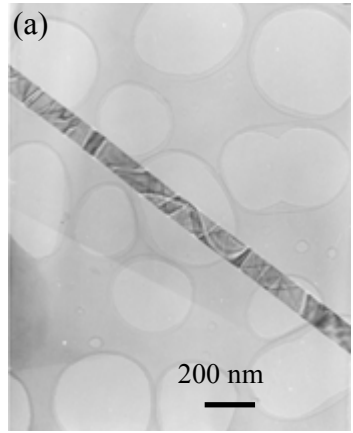 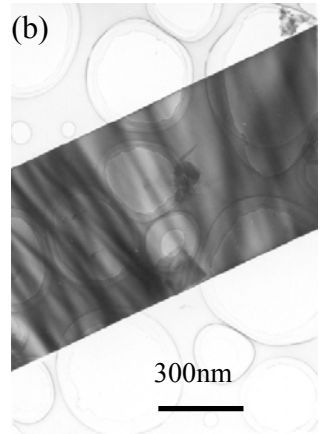

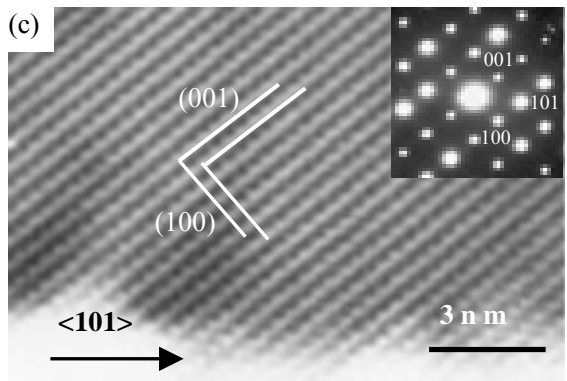 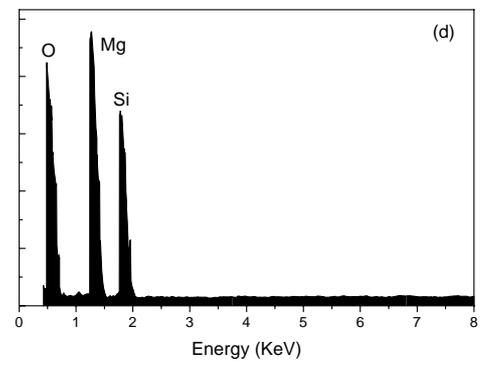

**Fig. 3 R. Zhu, et al.**

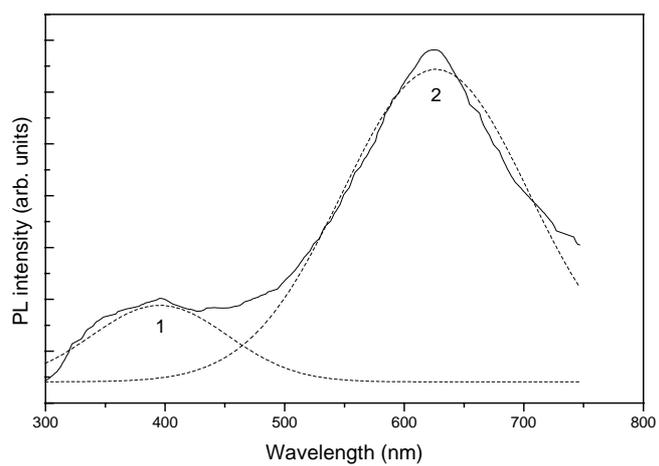

**Fig. 4 R. Zhu, et al.**

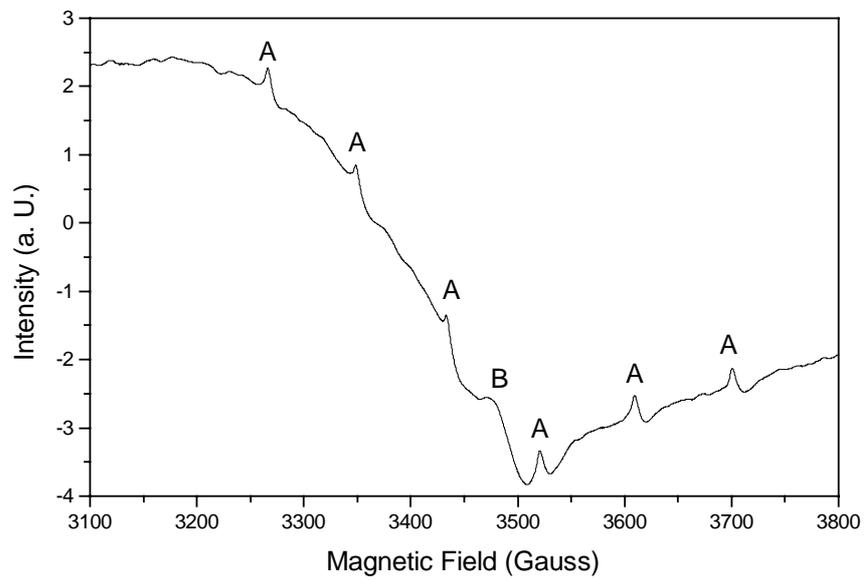

**Fig. 5 R. Zhu, et al.**